\documentclass[aps,pra,amsmath,amssymb,twocolumn,superscriptaddress,showpacs]{revtex4-1}
\usepackage{amsmath,amssymb}
\usepackage{graphicx}	
\usepackage{bm}

\def\vecb{\boldsymbol}

\begin{document}

\author{Shunsuke~A.~Sato}
\email{ssato@ccs.tsukuba.ac.jp}
\affiliation 
{Center for Computational Sciences, University of Tsukuba, Tsukuba 305-8577, Japan}
\affiliation 
{Max Planck Institute for the Structure and Dynamics of Matter, Luruper Chaussee 149, 22761 Hamburg, Germany}

\author{Angel~Rubio}
\affiliation 
{Max Planck Institute for the Structure and Dynamics of Matter, Luruper Chaussee 149, 22761 Hamburg, Germany}
\affiliation 
{Center for Computational Quantum Physics (CCQ), Flatiron Institute, 162 Fifth Avenue, New York, NY
10010, USA}

\title{Nonlinear electric conductivity and THz-induced charge transport in graphene}

\begin{abstract}
Based on the quantum master equation approach, the nonlinear electric conductivity of graphene is investigated under static electric fields for various chemical potential shifts. The simulation results show that, as the field strength increases, the effective conductivity is firstly suppressed, reflecting the depletion of effective carriers due to the large displacement in the Brillouin zone caused by the strong field. Then, as the field strength exceeds $1$~MV/m, the effective conductivity increases, overcoming the carrier depletion via the Landau--Zener tunneling process. Based on the nonlinear behavior of the conductivity, the charge transport induced by few-cycle THz pulses is studied to elucidate the ultrafast control of electric current in matter.
\end{abstract}

\maketitle

\section{Introduction \label{sec:intro}}

Recent developments of laser technologies have enabled the study of light-induced nonequilibrium electron dynamics in matter \cite{RevModPhys.72.545,RevModPhys.81.163,Schultze1348,Lucchini916,Zurch2017,Schmidt2018,Siegrist2019}. Among various intriguing materials, graphene has been attracting a large amount of interest owing to its unique electronic structure, the so-called Dirac cone. Various light-induced phenomena in graphene have been studied, such as the light-induced topological phase \cite{PhysRevB.79.081406}, high-order harmonic generation \cite{Mikhailov_2007,PhysRevB.82.201402,Yoshikawa736,Hafez2018,PhysRevB.103.L041408}, and light-induced anomalous Hall effects \cite{PhysRevB.79.081406,McIver2020,PhysRevB.99.214302}. Furthermore, nonlinear light-induced current injection in graphene has been investigated toward the achievement of ultrafast optoelectronic devices \cite{Higuchi2017,PhysRevLett.121.207401,Heide_2019}.

To control electric current and charge transport by light, a deep understanding of the field-induced nonequilibrium electron dynamics in matter is indispensable. The nonlinear conductivity of graphene in the THz regime has been investigated, and its field-induced transparency has been experimentally reported \cite{Hwang2013,Paul2013,Mics2015,doi:10.1063/1.4902999,Hafez2014,Choi2017}, reflecting the reduction in the conductivity caused by a strong field. Based on the semi-classical kinematic theory, the conductivity reduction has been explained through the change in the carrier scattering rate \cite{Bao2009,Bao2010}. By contrast, an enhancement of the conductivity has been suggested in a strong field regime via the Landau--Zener tunneling, which is based on the quantum nature of electrons in solids \cite{PhysRevB.81.165431}. Therefore, the microscopic physics behind the nonlinear conductivity of graphene still needs to be clarified via a full quantum mechanical description in order to provide a comprehensive understanding of its nonlinear opto-electronic properties.

In this work, the nonlinear electronic current in graphene is theoretically studied under strong static fields using the quantum master equation approach. As a result of the fully-quantum mechanical simulations, it is shown that the conductivity reduction in graphene can be understood in terms of the depletion of the effective carriers without considering the phenomenological change of the scattering rates. Furthermore, the depletion of carriers is found to be overcome by the Landau--Zener tunneling in the strong field regime, resulting in an enhancement of the effective conductivity as well as the induced current. On account of the highly nonlinear behavior of the conductivity, a method to induce charge transport via few-cycle THz laser pulses is proposed, providing a possible foundation for realizing ultrafast optoelectronics with graphene.

\section{Method}

Our calculation method has been described in detail elsewhere \cite{PhysRevB.99.214302,Sato2019,PhysRevB.103.L041408}, so we briefly describe the theoretical modeling here. To describe the light-indued electron dynamics in graphene, the following quantum master equation was employed \cite{PhysRevB.99.214302,Sato2019},
\begin{eqnarray}
\frac{d}{dt} \rho_{\vecb k}(t) = \frac{1}{i\hbar}\left [ \rho_{\vecb k}(t), H_{\vecb k}(t) \right ] + \hat D \left [\rho_{\vecb k}(t) \right ],
\end{eqnarray}
where $\rho_{\vecb k}(t)$ is the one-body reduced density matrix at each $\vecb k$-point, $\hat D\left [\rho_{\vecb k}(t) \right ]$ is the relaxation operator, and $H_{\vecb k}(t)$ is the Hamiltonian. The simple two-band approximation was employed for the electronic structure of graphene as
\begin{eqnarray}
H_{\vecb k}(t) = v_\mathrm{F} \sigma_x \left [k_x +A_x(t) \right ]
+v_\mathrm{F} \sigma_y \left [ k_y +A_y(t)\right ],
\label{eq:ham}
\end{eqnarray}
where $\sigma_j$ are the Pauli matrices, $k_j$ are the $j$th components of the Bloch wave vector $\vecb k$, and $A_j(t)$ is the $j$ component of the vector potential $\vecb A(t)$, which is related to the external electric field, according to $\vecb E(t) = -\dot {\vecb A}(t)$ in the dipole approximation. The Fermi velocity $v_\mathrm{F}$ was set to $1.12\times 10^6$~m/s in accordance with a previous \textit{ab initio} calculation \cite{PhysRevLett.101.226405}. The relaxation operator $\hat D\left [\rho_{\vecb k}(t) \right ]$ was constructed using the relaxation time approximation \cite{PhysRevLett.73.902}. $\hat D\left [\rho_{\vecb k}(t) \right ]$ depends on the longitudinal relaxation time $T_1$, transverse relaxation time $T_2$, electron temperature $T_e$, and chemical potential $\mu$. $T_1$, $T_2$, and $T_e$ were set to $100$~fs, $20$~fs, and $300$~K, respectively, whereas $\mu$ was treated as a tunable parameter. Our treatment of the relaxation operator has been described in detail elsewhere \cite{PhysRevB.99.214302,Sato2019,PhysRevB.103.L041408}. As will be shown later (see Figs.~\ref{fig:graphene_dc_sigma_intra}~(c--f)), we consider the electron dynamics only in a narrow region around the Dirac point. Therefore, the linear band approximation of Eq.~(\ref{eq:ham}) is expected to be valid.

The electron dynamics in graphene was computed under a static electric field described as $\vecb A(t) = -E_0 \vecb e_x t$, where $E_0$ is the strength of the applied field, and $\vecb e_x$ is the unit vector along the $x$-direction. After sufficient time has elapsed, the system reaches a nonequilibrium steady state due to the balance between the field-induced excitation and intrinsic relaxation. The electric current in the nonequilibrium steady state can be evaluated as
\begin{eqnarray}
\vecb J(E_0) = \lim_{t\rightarrow \infty} \frac{(-1)}{(2\pi)^2}\int d\vecb k \mathrm{Tr}
\left [\rho_{\vecb k}(t) \frac{\partial H_{\vecb k}(t)}{\partial \vecb A(t)} \right ].
\label{eq:current}
\end{eqnarray}

Due to the circular symmetry of the Hamiltonian in Eq.~(\ref{eq:ham}), the induced current has only the $x$-component, i.e., $\vecb J(E_0)=J(E_0) \vecb e_x$. With this notation, the effective conductivity is here denoted as $\sigma(E_0) = J(E_0)/E_0$. It should be noted that, in the weak field limit, the effective conductivity $\sigma(E_0)$ approaches the linear conductivity i.e., $\sigma_{0}=\lim_{E_0 \rightarrow 0}\sigma(E_0)$.

\section{Results}

\begin{figure}[htbp]
  \centering
  \includegraphics[width=0.90\columnwidth]{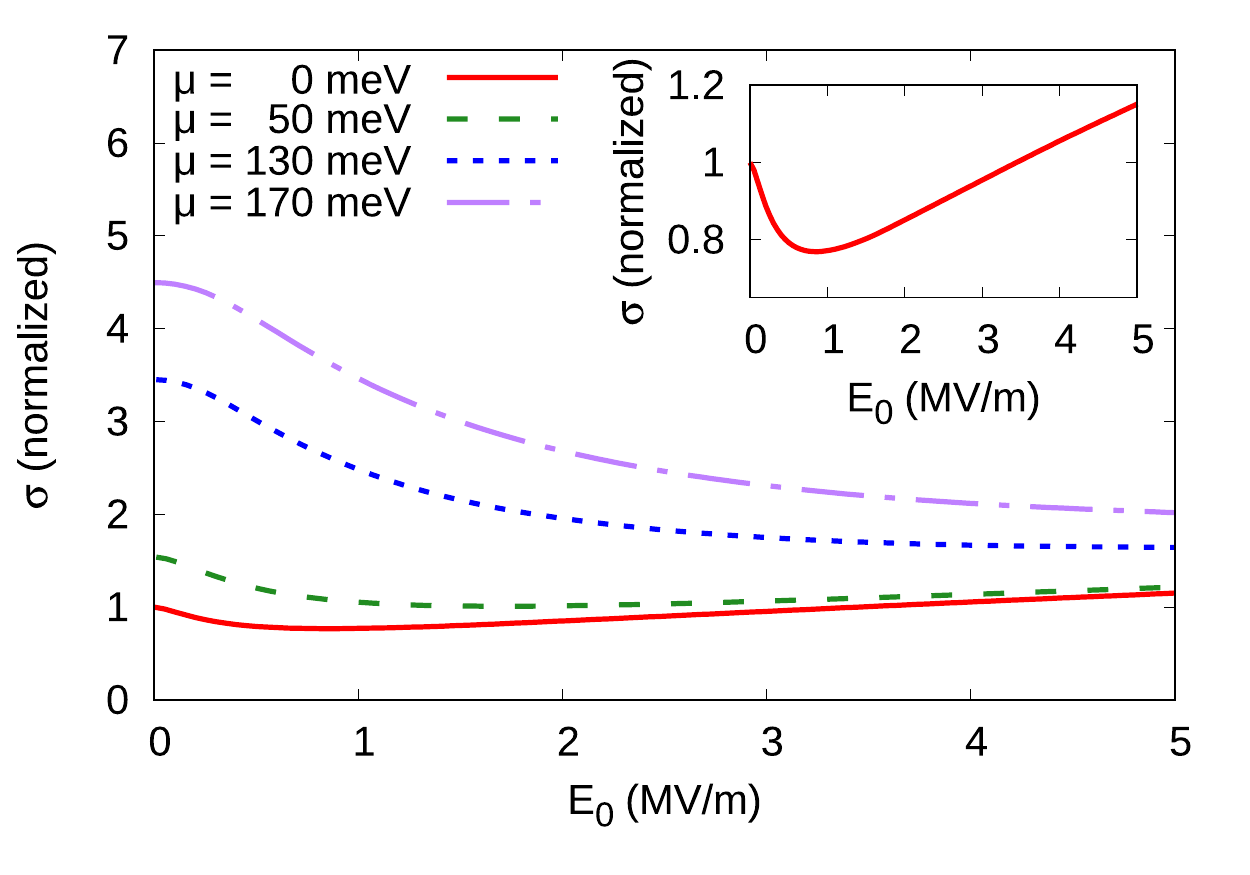}
\caption{\label{fig:graphene_dc_sigma}
Nonlinear conductivity of graphene as a function of the field strength $E_0$ for different values of the chemical potential $\mu$. The results for $\mu=0$ are shown in the inset. All the results are normalized by the linear conductivity $\sigma_0$ with $\mu=0$.
}
\end{figure}

Assuming that the THz fields vary slowly in time and the induced electron dynamics is well approximated by the quasi-static description, the THz and dc conductivity of graphene were investigated based on the effective conductivity $\sigma(E_0)$. This approximation becomes accurate when the THz field frequency $\omega_{\mathrm{THz}}$ is sufficiently smaller than the intrinsic relaxation rates such as $1/T_1$ and $1/T_2$. Figure~\ref{fig:graphene_dc_sigma} shows the computed conductivity $\sigma(E_0)$ as a function of the field strength $E_0$ for different values of the chemical potential $\mu$. In the relatively weak field regime, the conductivity decreases upon increasing the applied field strength for all the investigated chemical potentials. Since the energy loss of the external field is provided by Joule heating, namely, $J(E_0)E_0 = \sigma(E_0)E^2_0$, the conductivity reduction is directly related to the field-induced transparency of graphene observed in the experiments \cite{Hwang2013,Paul2013,Mics2015,doi:10.1063/1.4902999,Hafez2014,Choi2017}. Using a semi-classical model, this conductivity reduction has been previously explained in terms of an enhancement of the carrier scattering rate \cite{Bao2009,Bao2010}. However, the microscopic mechanism of this conductivity reduction has not been yet clarified on the basis of a quantum description. Remarkably, the proposed fully-quantum model can describe the reduction of the electric conductivity without considering the change of the relaxation times, $T_1$ and $T_2$. Hence, these results indicate the existence of yet another microscopic mechanism behind the field-induced transparency of graphene, which is different from the change in the scattering rate observed in the classical description.

The inset of Fig.~\ref{fig:graphene_dc_sigma} illustrates the nonlinear conductivity $\sigma(E_0)$ at the charge neutrality point ($\mu=0$). While the conductivity decreases as the field strength increases up to around $E_0=1$~MV/m, the conductivity starts to increase almost linearly. In a previous work, an increase in the conductivity via the Landau--Zener tunnel mechanism was suggested \cite{PhysRevB.81.165431}. However, this enhancement mechanism has not been investigated in combination with relaxation effects. Therefore, the interplay between the tunneling mechanism and relaxation needs to be considered in order to develop a comprehensive understanding of the nonlinear carrier transport in graphene.

To obtain further insight into these phenomena, the contribution from the intraband current was evaluated. For this purpose, the instantaneous eigenstates were defined as $H_{\vecb k}(t)\left |u_{b,\vecb k+\vecb A(t)} \right \rangle = \epsilon_{b,\vecb k+\vecb A(t)} \left |u_{b,\vecb k+\vecb A(t)} \right \rangle$, where $b$ denotes the band index, i.e., valence $(b=v)$ or conduction ($b=c$) bands. The intraband current is then defined as
\begin{eqnarray}
\vecb J^{\mathrm{intra}}(E_0) = \sum_{b=v,c}\lim_{t\rightarrow \infty} \frac{(-1)}{(2\pi)^2}\int d\vecb k \frac{\partial \epsilon_{b,\vecb k + \vecb A(t)}}{\partial \vecb k} n_{b,\vecb k+\vecb A(t)}, \nonumber \\
\label{eq:current-intra}
\end{eqnarray}
where the band population $n_{b,\vecb k+\vecb A(t)}$ is defined as $n_{b,\vecb k + \vecb A(t)}=\langle u_{b,\vecb k + \vecb A(t)}|\rho_{\vecb k}(t)| u_{b,\vecb k + \vecb A(t)}\rangle$.

\begin{figure}[htbp]
  \centering
  \includegraphics[width=0.90\columnwidth]{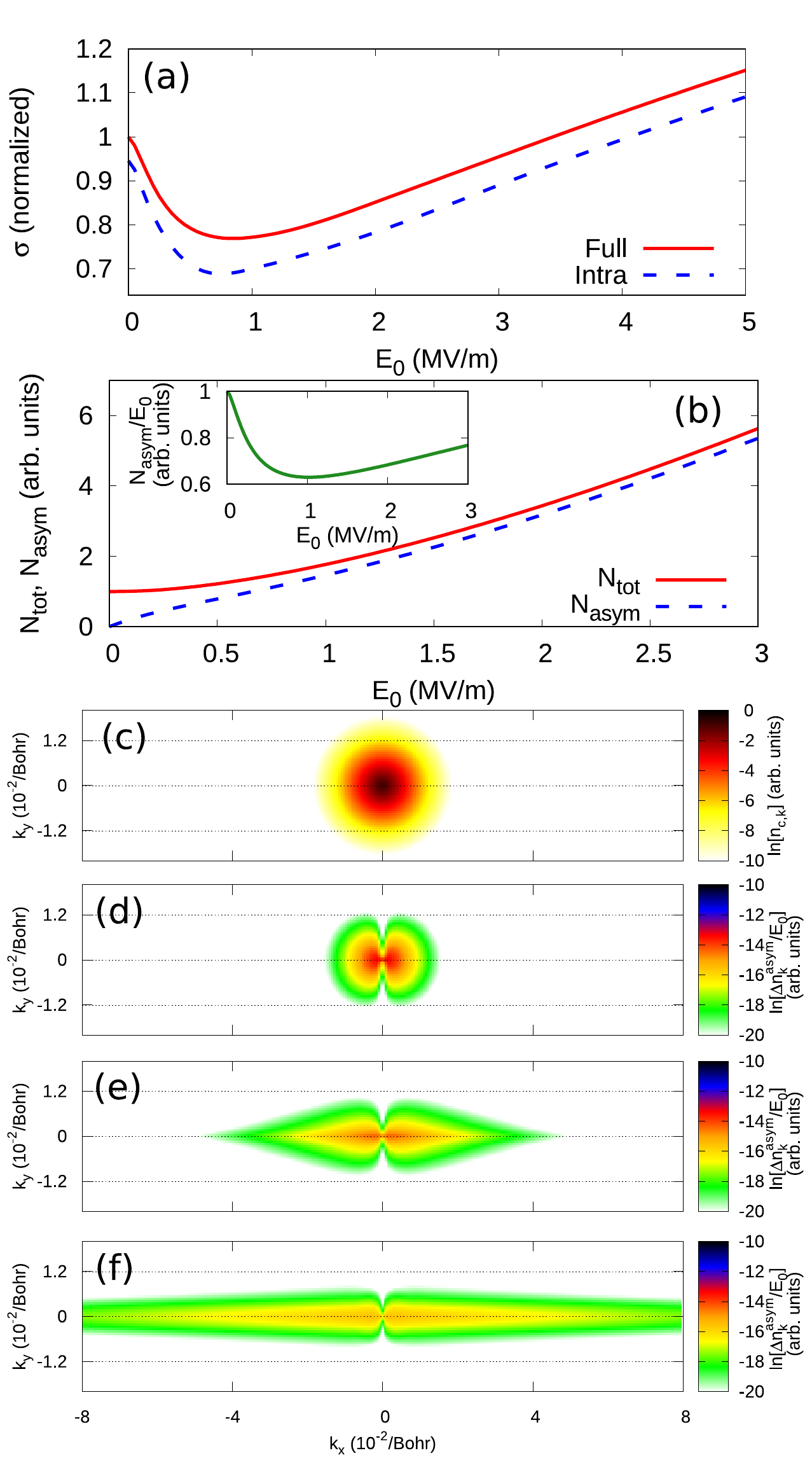}
\caption{\label{fig:graphene_dc_sigma_intra}
(a) Effective conductivities evaluated with the full current (red solid line) and intraband current (blue dashed line). (b) Conduction population $N_{tot}$ (red solid line) and asymmetric population $N_{asym}$ (blue dashed line) as a function of the field strength $E_0$. The asymmetric population normalized by the field, $N_{asym}/E_0$, is shown in the inset. (c) Population distribution in the conduction band, $n_{c}(\vec k)$, in the absence of the field. (d--f) Induced asymmetric population $n_{asym}(\vecb k)$ around the Dirac point for different values of the field strength $E_0$; (d) $10^{-3}$~MV/m, (e) $1$~MV/m, and (f) $5$~MV/m. All quantities in the figure are computed for $\mu=0$.
}
\end{figure}

Figure~\ref{fig:graphene_dc_sigma_intra}~(a) shows the conductivities computed using the full current $\vecb J(E_0)$ and intraband current $\vecb J^{\mathrm{intra}}(E_0)$. The contribution from the intraband current dominates the total conductivity in the whole range of field strengths investigated. Therefore, the band velocity $\partial \epsilon_{b,\vecb k}/\partial \vecb k$ and band population $n_{b,\vecb k}$ play a central role in the nonlinear conductance of graphene.

Due to the circular symmetry of the model, the band velocity exhibits a circular symmetry. Therefore, the intraband current in Eq.~(\ref{eq:current-intra}) originates from the non-symmetric population distribution in the nonequilibrium steady state. To elucidate the population distribution under field, the total conduction population, $N_{tot}  =  \int d\vecb k n_{c,\vecb k}$, and corresponding asymmetric population, $N_{asym} = \int d\vecb k \left | n_{c,\vecb k}-n_{c,-\vecb k} \right |/2$ were evaluated. Figure~\ref{fig:graphene_dc_sigma_intra}~(b) shows the total and asymmetric populations as a function of the field strength $E_0$ for $\mu=0$. In the zero field limit, the total population $N_{tot}$ reaches a finite value due to thermally excited carriers, whereas the asymmetric population $N_{asym}$ vanishes. In the weak field regime (below $E_0=1$~MV/m), the asymmetric population shows a faster increase than the total population $N_{tot}$ upon increasing the field strength. However, $N_{asym}$ is always smaller than $N_{tot}$ by definition, resulting in a slowdown of the increase in $N_{asym}$. As a result, the field-normalized asymmetric population $N_{asym}/E_0$ (see the inset of Fig.~\ref{fig:graphene_dc_sigma_intra}~(b)) is characterized by a similar reduction to that of the conductivity shown in Fig.~\ref{fig:graphene_dc_sigma_intra}~(a). Hence, the conductivity reduction can be understood in terms of the suppression of the field-induced asymmetric population. The suppression of the asymmetric distribution can be further understood based on the depletion of the thermal carriers: in the weak field regime, the field-induced asymmetric distribution is formed as a consequence of the displacement of the thermal carriers caused by the applied field. However, once the field strength becomes relatively large, a large proportion of the thermal carriers are already displaced, and a larger asymmetric population cannot be formed. As a result, the effective conductivity is suppressed due to the suppression of the asymmetric distribution. The same mechanism for the doped carriers should explain the conductivity reduction for the finite chemical potential shifts.

In contrast to the weak and moderate field regimes, the total conduction population $N_{tot}$ increases significantly upon increasing the field strength in the strong field regime ($E_0>1$~MV/m), as shown in Fig.~\ref{fig:graphene_dc_sigma_intra}~(b). This can be understood based on the carrier generation through the Landau--Zener tunneling. Consequently, the depletion of the conduction population is eliminated, and the asymmetric population $N_{asym}$ increases significantly, resulting in the increase of the effective conductivity in the strong field regime. Therefore, the non-monotonic behavior of the effective conductivity of graphene can be understood through the competition between the depletion of the effective carriers and the generation of additional carriers via the Landau--Zener tunneling.

To develop further microscopic insight into the nonlinear conductivity of graphene under strong fields, the conduction population distribution $n_{c,\vecb k}$ in the $\vecb k$-space was evaluated. Figure~\ref{fig:graphene_dc_sigma_intra}~(c) shows the conduction band population $n_{c,\vecb k}$ at $T_e=300$~K in the absence of a static electric field. Here, the Dirac point is set at the origin of the coordinates. As can be seen from Fig.~\ref{fig:graphene_dc_sigma_intra}~(c), the thermally excited conduction carriers exhibit a circular distribution as a consequence of the circular symmetry of the Hamiltonian. Once a static electric field is applied, the field breaks the circular symmetry of the population distribution, resulting in the intraband current.

To study the field-induced symmetry breaking, the asymmetric population distribution $\Delta n^{\mathrm{asym}}_{\vecb k}= \left |n_{c,\vecb k}-n_{c,-\vecb k} \right |/2$ was evaluated under a static field. Figures~\ref{fig:graphene_dc_sigma_intra}~(d--f) show the field-normalized asymmetric distribution $\Delta n^{\mathrm{asym}}_{\vecb k}/E_0$ for different field strengths. As can be seen from Fig.~\ref{fig:graphene_dc_sigma_intra}~(d), the induced distribution $\Delta n^{\mathrm{asym}}_{b,\vecb k}$ is close to the circular symmetry distribution in the weak field regime ($E_0=10^{-3}$~MV/m), reflecting the circular symmetry distribution of the thermal carriers. By contrast, once the field strength increases, the induced asymmetric distribution is elongated along the field direction, as shown in Figs.~\ref{fig:graphene_dc_sigma_intra}~(d) and (e). The elongation can be understood through the intraband motion of the carriers: since the field is aligned along the $x$-axis, the carriers move along this axis, and a significant asymmetric distribution is thus formed. Once the field strength becomes even larger, the carriers can move across larger distances in the $\vecb k$-space before being scattered. As a result, a significant elongation along the field direction can be formed.

\begin{figure}[htbp]
  \centering
  \includegraphics[width=0.90\columnwidth]{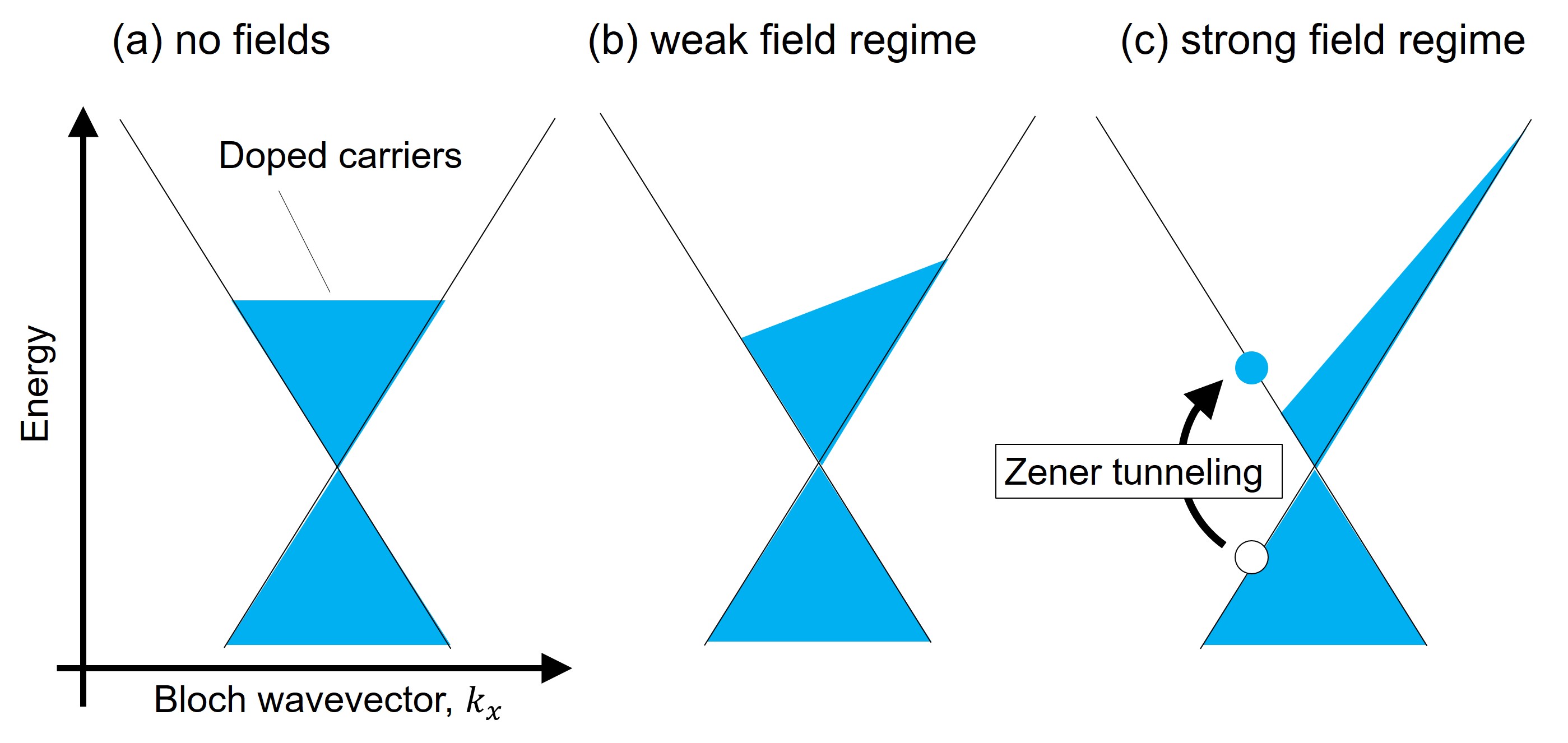}
\caption{\label{fig:schematics}
Schematic picture of microscopic mechanism of the nonlinear conductivity of graphene: the electron distribution in the Dirac cone (a) in the absence of the field, (b) under a weak field, and (c) under a strong field.
}
\end{figure}

We schematically summarize the microscopic picture behind the nonlinear conductivity of graphene in Fig.~\ref{fig:schematics}. For simplicity, we consider the electron-doped system here. Figure~\ref{fig:schematics}~(a) shows the electron distribution in the Dirac cone in the absence of the field. Once a weak field is applied, the electron distribution is displaced in the Brillouin zone, as shown in Fig.~\ref{fig:schematics}~(b), resulting in the intraband current. If the applied field strength becomes stronger, a large proportion of the carriers are displaced, as shown in Fig.~\ref{fig:schematics}~(c), causing the depletion of carriers and suppression of the conductivity. When the field strength becomes even stronger, the Landau--Zener process is activated, and additional carriers are supplied from the valence bands via the excitation, as described in Fig.~\ref{fig:schematics}~(c). As a result, the carrier depletion is overcome, and the nonlinear conductivity starts increasing.

\begin{figure}[htbp]
  \centering
  \includegraphics[width=0.90\columnwidth]{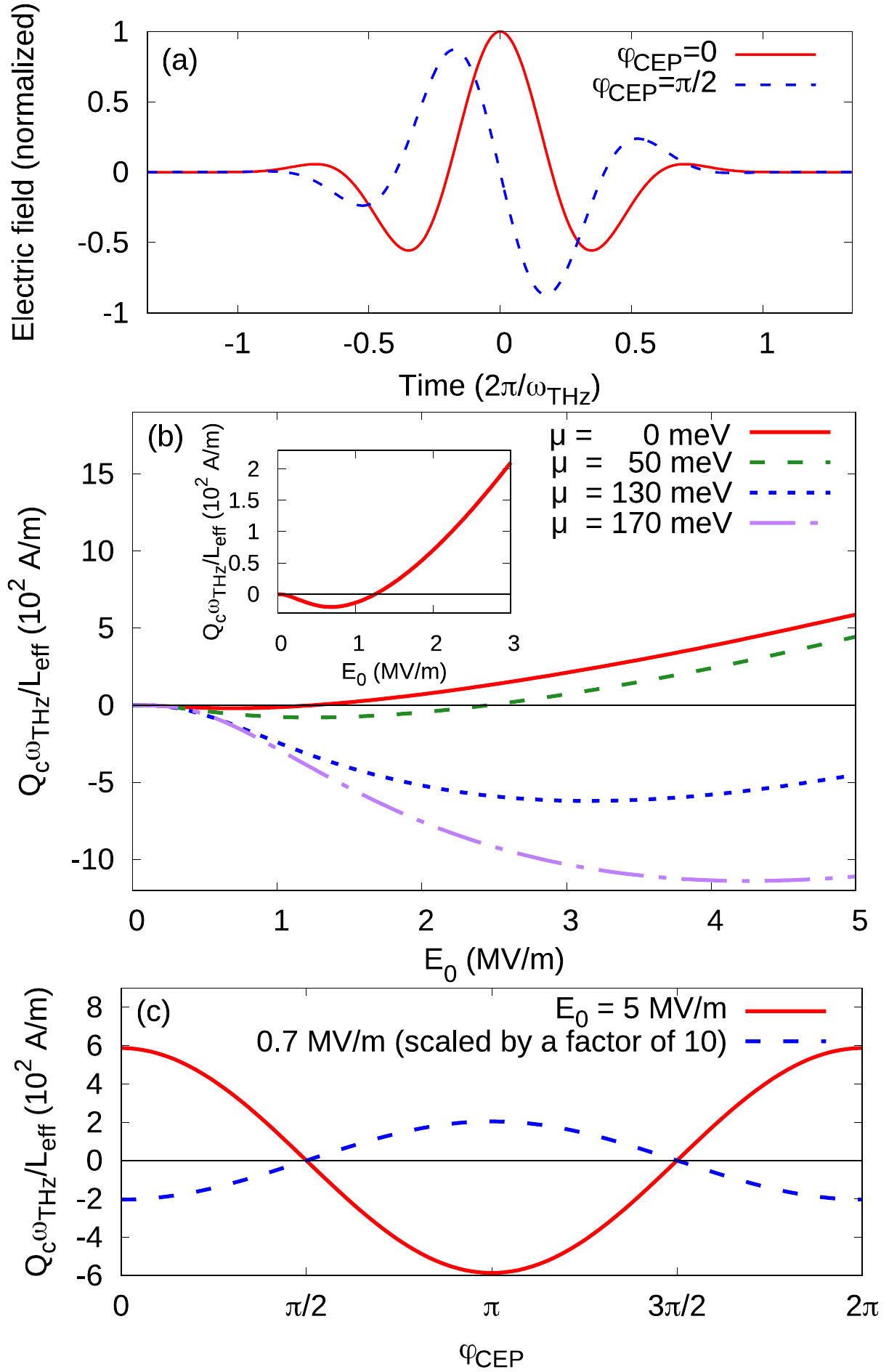}
\caption{\label{fig:charge_fig}
(a)~Time profiles of the applied THz field with different CEPs. (b)~Charge transported by the THz pulses as a function of the peak field strength for different values of the chemical potentials $\mu$. The case for $\mu=0$ is shown in the inset. (c)~CEP dependence of the transported charge.
}
\end{figure}

Having clarified the microscopic physics behind the nonlinear conductivity of graphene, a method is here proposed to control the charge transport in graphene via few-cycle THz pulses. For a THz pulse, the following form of the electric field was considered in the domain $-T_\mathrm{d}/2<t<T_\mathrm{d}/2$,
\begin{eqnarray}
E(t) = \frac{E_0}{\omega_{\mathrm{THz}}} \frac{d}{dt} \left [- \sin \left (\omega_{\mathrm{TH}z} t + \phi_{\mathrm{CEP}}  \right )\cos^8 \left (\pi \frac{t}{T_\mathrm{d}} \right )  \right ], \nonumber \\
\end{eqnarray}
whereas the field was set to zero outside this domain. Here, $\omega_{\mathrm{THz}}$ is the mean frequency of the THz pulse, and $\phi_{\mathrm{CEP}}$ is the carrier--envelope phase (CEP). The pulse duration $T_\mathrm{d}$ is set to $5.38\pi/\omega_{\mathrm{THz}}$ so that the full width at half maximum of the pulse becomes half of the cycle, i.e., $\pi/\omega_{\mathrm{THz}}$. Figure~\ref{fig:charge_fig}~(a) shows the time profile of the applied electric fields with different CEP values ($\phi_{\mathrm{CEP}}=0, \pi/2$). As can be seen from the figure, the few-cycle pulses can give rise to inversion symmetry breaking, depending on the value of $\phi_{\mathrm{CEP}}$.

Based on the inversion symmetry breaking of few-cycle light pulses, light-induced charge transport can be realized \cite{Schiffrin2013,PhysRevLett.113.087401,Higuchi2017}. To evaluate the transported charge, the THz fields were assumed to vary sufficiently slowly, so that the induced current could be determined via the instantaneous field strength as $J(t)=J\left ( E(t) \right )$. The transported charge is then defined by means of the time integration of the current $Q_c = L_{\mathrm{eff}} \int^{\infty}_{-\infty}dt J\left (E(t) \right )$, where $L_{\mathrm{eff}}$ is the cross-length that corresponds to the cross-section for bulk systems \cite{Schiffrin2013,PhysRevLett.113.087401}. It should be noted that the transported charge is proportional to the inverse frequency, $1/\omega_{\mathrm{THz}}$, because of the instantaneous field dependence of the current, $J(t)=J\left ( E(t) \right )$. Although the amount of the transported charge $Q_c$ depends on the effective cross-length $L_{\mathrm{eff}}$ and frequency $\omega_{\mathrm{THz}}$, one can introduce a normalized quantity as $Q_c \omega_{\mathrm{THz}}/L_{\mathrm{eff}}$, which is invariant to $L_{\mathrm{eff}}$ and $\omega_{\mathrm{THz}}$.

Figure~\ref{fig:charge_fig}~(b) shows the noamlized transported-charge $Q_c \omega_{\mathrm{THz}}/L_{\mathrm{eff}}$ as a function of the peak field strength $E_0$ of the THz pulses for $\phi_\mathrm{CEP}=0$ and different chemical potential values. By increasing the peak field strength, the charge is first transported in the direction opposite to the peak field direction for all the investigated chemical potentials. A larger amount of charge is transported for a larger chemical potential shift $\mu$. These features of the light-induced charge transport can be understood in terms of the suppression of the nonlinear conductivity $\sigma(E_0)$ shown in Fig.~\ref{fig:graphene_dc_sigma}. This figure shows that the nonlinear conductivity first decreases upon increasing the field strength. Thus, the induced current around the pulse peak timing is suppressed. As a result, the charge transport toward the field peak direction is also suppressed, and the total charge is effectively transported toward the opposite direction. Furthermore, since a larger reduction in the conductivity is observed for the largest chemical potential shift investigated, the resulting charge transport becomes also larger for higher $\mu$ values.

The inset of Fig.~\ref{fig:charge_fig}~(b) displays the transported charge for $\mu=0$. Here, the charge is first transported in the direction opposite to the peak field direction. The transport direction then switches once the peak field strength exceeds $1.25$~MV/m. The change in direction of the transported charge originates from the change in the trend of the nonlinear conductivity: in the relatively weak field regime, the effective conductivity for $\mu=0$ decreases upon increasing the field strength due to the depletion of the thermal carriers, resulting in the observed negative charge transport driven by the few-cycle THz pulses. By contrast, in the strong field regime, the conductivity increases with the increase of $E_0$ due to the carrier generation though the Landau--Zener tunneling, thus resulting in the positive charge transport. Hence, the direction of the charge transport can be controlled by tuning the field strength. It should be noted that a similar sign change in charge transport was observed in the infrared field regime and was interpreted in terms of the interference of photo-injected carriers \cite{Higuchi2017,PhysRevLett.121.207401,Heide_2019}. By contrast, the results of the present work are based on the quasi-static picture, and thus the microscopic mechanism behind the present results is different from the one proposed in previous works.

The CEP dependence of the transported charge was investigated further. Figure~\ref{fig:charge_fig}~(c) shows the transported charge as a function of $\phi_{\mathrm{CEP}}$. Here, $\mu$ is set to zero, and the results for the different peak field strength are shown. As can be seen from the figure, the magnitude and sign of the transported charge can be controlled by manipulating the CEP in both the depletion ($E_0=0.7$~MV/m) and tunneling ($E_0=5$~MV/m) regimes. The results indicate that ultrafast control of the electric current by means of CEP tuning can be realized also in the THz regime, similar to what has been demonstrated for the infrared regime \cite{Schiffrin2013,Higuchi2017}.

\section{Conclusion}

In conclusion, the nonlinear conductivity of graphene was investigated using a fully-quantum mechanical model, and it was shown that the effective conductivity exhibits a non-monotonic behavior as a function of the field strength. In the relatively weak field regime, the conductivity decreases upon increasing the field strength, reflecting the depletion of the effective carriers. By contrast, in the strong field regime, the conductivity increases as the field strength due to the carrier generation though the Landau--Zener process. Based on these findings, the possibility of controlling the electric current with few-cycle THz pulses was explored. It was demonstrated that the magnitude and sign of the transported charge can be controlled by manipulating the peak field strength $E_0$, the CEP $\phi_{\mathrm{CEP}}$, and the chemical potential shift $\mu$. These findings may open paths toward achieving ultrafast control of charge transport by light through nonequilibrium and nonlinear electron dynamics in matter.

\begin{acknowledgments}
This work was supported by JSPS KAKENHI Grant Number JP20K14382, the European Research Council (ERC-2015-AdG694097), the Cluster of Excellence 'Advanced Imaging of Matter' (AIM), Grupos Consolidados (IT1249-19) and SFB925 "Light induced dynamics and control of correlated quantum systems". We thank Enago for the English language review. The Flatiron Institute is a division of the Simons Foundation.
\end{acknowledgments}
\bibliography{ref}

\end{document}